\documentclass{article}

\usepackage{amssymb}
\usepackage{bm}
\newcommand\be{\begin{equation}}
\newcommand\ee{\end{equation}}
\newcommand\no{\nonumber\\}
\newcommand\beqa{\begin{eqnarray}}
\newcommand\eeqa{\end{eqnarray}}
\newcommand\bea{\begin{eqnarray*}}
\newcommand\eea{\end{eqnarray*}}
\newcommand\tet{\theta}
\newcommand\la{\lambda}
\newcommand{\lb}{\label}
\newcommand\al{\alpha}
\newcommand\bet{\beta}
\newcommand\ga{\gamma}
\newcommand\de{\delta}
\newcommand\s{\sigma}
\newcommand\e{\epsilon}

\newcommand\A{{\cal A}}
\newcommand\B{{\cal B}}

\newcommand\La{{\cal L}}

\newcommand\fr{\frac}
\newcommand\del{\partial}

\newcommand\ti{\tilde}

\newcommand\lp{\left(}
\newcommand\rp{\right)}
\newcommand{\eij}{\epsilon_{ij}}
\newcommand{\eabg}{\epsilon_{\al\beta \gamma}}
\newcommand{\easg}{\epsilon_{\al\sigma \gamma}}
\newcommand{\ebsg}{\epsilon_{\beta\sigma \gamma}}

\begin{document}
\begin{center}

{\LARGE Dynamics of Dipoles and  Quantum Phases  in Noncommutative Coordinates  }

\vspace{15mm}

\"{O}mer F. Dayi\footnote{E-mail addresses:
dayi@gursey.gov.tr and dayi@itu.edu.tr.}

\vspace{5mm}

\noindent{\it Physics Department, Faculty of Science and
Letters, Istanbul Technical University, TR-34469 Maslak--Istanbul, Turkey,}

and

\noindent {\it  Feza G\"{u}rsey Institute, P.O. Box 6,
TR--34684, \c{C}engelk\"{o}y--Istanbul, Turkey. }

\end{center}

\vspace{1cm}

{\small  The dynamics of a spin--1/2 neutral particle possessing   electric
and magnetic dipole moments interacting with
external electric and magnetic fields in noncommutative coordinates is obtained.
Noncommutativity of  space is interposed in terms of
a semiclassical constrained Hamiltonian system.
The relation between the quantum phase acquired by a particle interacting with an
electromagnetic field and
the (semi)classical force acting on the system is examined and generalized to
 establish a  formulation of the quantum
phases in noncommutative coordinates. The general  formalism is  applied to
physical systems yielding the
Aharonov--Bohm, Aharonov--Casher, He--McKellar--Wilkens and Anandan phases
in noncommutative coordinates.
Bounds for the noncommutativity parameter $\theta$ are derived
comparing the deformed phases with the experimental data on
the Aharonov--Bohm and Aharonov--Casher phases.    }

\vspace{1cm}

\section{Introduction}

The wave function of a particle encircling  the sources of
electromagnetic fields may acquire a  phase factor which cannot be removed by gauge or phase transformations.
The first predicted quantum phase is the well--established
Aharonov--Bohm (AB)
phase\cite{ab}. Later for spin--1/2
 neutral particles possessing  magnetic dipole moment a similar effect was predicted
by Aharonov--Casher (AC)\cite{ac}.
Although, it has not been experimentally verified yet
He--McKellar and Wilkens (HMW) derived
the dual of the AC phase for neutral
particles with an electric dipole moment\cite{hmw}.
In  \cite{ana1} Anandan studied the dynamics of a
spin--1/2 neutral particle possessing both electric and magnetic dipole moments interacting with
external electric and magnetic fields which leads to a quantum phase for a specific
configuration\cite{flr}.
These quantum phases have already been studied in  noncommutative space.
For the AB phase in noncommutative coordinates there appeared different results
depending on  the velocity \cite{cpst,glr} and independent of the velocity \cite{ao}.
However, for the AC \cite{mz,lw},
HMW \cite{wl} and the Anandan \cite{prfn} phases in noncommutative space
similar results which explicitly
depend on  eigenvalues of the momentum operator $\hat p_\mu ,$
are obtained.
In  all these studies noncommutativity of the coordinates  $r_\mu$
is  mainly imposed by  shifting   the coordinates as
$
r_\mu \rightarrow r_\mu -\fr{i}{2\hbar}\theta_{\mu\nu}\hat p^\nu ,
$
where $\theta$ is  the constant noncommutativity parameter.
Within this approach one attains a $\tet$ deformed Schroedinger equation or equivalently a
$\tet$ deformed Hamiltonian which are then
employed to calculate the related quantum phases. Obviously, this scheme is very
sensitive how the  external fields depend on the coordinates $r_\mu .$

We are concerned with deriving the dynamical equations of a spin--1/2 particle possessing
  dipole moments interacting with  electromagnetic fields in noncommutative coordinates as well as
obtaining the $\theta$ deformed quantum phases
 in the framework of the
semiclassical constrained Hamiltonian formalism proposed in \cite{om}.
This semiclassical approach which was invented
to study the Berry gauge fields\cite{ber} arising in spin dynamics,
is suitable to study gauge systems in terms of field strengths without referring to the
explicit
realization of the gauge fields.
Moreover, it is adequate to study spin--dependent dynamical systems
in noncommutative coordinates as was shown in \cite{mo}.
We briefly review the formalism of \cite{om} in the next section.
In  the third section
we first present  the general formulation of
deriving the equations of motion and  semiclassical
force in noncommutative coordinates
and then  apply it to
a spin--1/2 neutral particle  possessing  electric and magnetic
dipole moments interacting
with  external
electric and magnetic fields in noncommutative coordinates.

Our semiclassical  constrained Hamiltonian  formalism
does not give rise to an effective Hamiltonian which can be
utilized to discern the  quantum phases
in noncommutative coordinates. Instead, we
are equipped with the equations of motion and  the force acting on the system.
Fortunately, in \cite{spa} it was shown that the quantum phase acquired by a particle
interacting with an electromagnetic field is related to the classical force acting on the
system.
By generalizing this relation to noncommutative systems we will present a definition of
the quantum phase in  noncommutative space,
which is velocity independent. We apply the formulation
to obtain the AB phase and the Anandan  phase in noncommutative
coordinates.
Obviously, from the Anandan phase  the AC and HMW phases
follow, respectively, for vanishing electric and magnetic dipole moments.
We then argue that our formulation can be generalized to formulate the quantum phase
attained by a particle interacting with the external gauge field taking values
in the noncommutative $U(N).$
In the last
section we compare the $\tet$ deformed AB  and AC phases which we obtained
with the experimental data to derive  bounds on
the  noncommutativity parameter $\tet .$

\section{The semiclassical approach}
\label{sca}

We would like to present briefly
the semiclassical scheme established in \cite{om} for studying spin dependent dynamical
systems. We consider  matrix observables which
are functions of the  classical phase space variables $(\pi_\mu , x_\mu ),$
though they are allowed to be
$\hbar $ dependent.
One of the main ingredients is   the semiclassical bracket of
the  observables  $K(\pi ,x)$ and $N(\pi ,x)$  defined as
\be
 \{K(\pi,x ) , N(\pi,x ) \}_C \equiv
\frac{-i}{\hbar}[K,N]+ \no
\frac{1}{2} \{ K(\pi,x ) , N(\pi,x ) \}
-\frac{1}{2} \{ N(\pi,x ) , K(\pi,x ) \} , \lb{SCB}
\ee
which is obtained from
the Moyal bracket  by retaining the lowest two terms in $\hbar$.
The first term  is the ordinary commutator of matrices, it is not the quantum commutator.
The others are the  Poisson brackets defined as
$$
 \{ K(\pi,x ) , N(\pi,x ) \} \equiv
 \frac{ \partial K}{ \partial x^\nu}\fr{ \partial N}{ \partial \pi_\nu}
-\fr{\partial K }{ \partial \pi_\nu }\fr{\partial N}{ \partial x^\nu} .
$$
$K$ and $N$ may depend on $\hbar ,$ but only the two lowest order terms
in $\hbar$ are detained in (\ref{SCB}), so that
it is a semiclassical approximation.
The semiclassical bracket (\ref{SCB})
and the
semiclassical Hamiltonian  $H(\pi ,x)$ are employed
to define
\be
{\dot K } (\pi,x) = \{K(\pi,x ) , H(\pi,x ) \}_C ,\lb{evo}
\ee
where the dot
over the observables indicates the time
derivative.
This designates the semiclassical  dynamics.

\subsection{A semiclassical constrained Hamiltonian system}

To generate  the semiclassical dynamics one
substitutes the Poisson brackets
with the semiclassical brackets (\ref{SCB})
in the ordinary Hamiltonian dynamical relations.
To achieve   a systematic method of
describing the dynamics in  noncommutative space,
we introduce the  first--order matrix Lagrangian
$$
\La   = {\dot r}^\al [ I p_\al /2 +\rho \A_\al (r,p) ]
-{\dot p}^\al [ I r_\al /2 -\xi \B_\al (r,p) ] - H_0(r,p), \lb{oL}
$$
in terms of the gauge fields
$\A_\al , \B_\al $ which are in general  matrix valued.
$\rho , \xi   $
are the coupling constants and
$I$ denotes the unit matrix.
In this work we only deal with  three dimensional  systems, so that $\al,\beta =1,2,3. $
This  Lagrangian   leads to  some constraints which are second
class. To effectively set them to zero
we employ the Dirac procedure by  substituting
the semiclassical bracket of observables (\ref{SCB})  with
the semiclassical Dirac bracket defined adequately. In fact one can calculate that
the phase space variables  satisfy the following semiclassical Dirac brackets
\beqa
\{r^\al,r^\beta \}_{CD}  & = &
\xi  G^{\al\beta  }
-\rho\xi^2(GFG)^{\al\beta}+\cdots
,\lb{rr1}\\
\{p^\al,p^\beta \}_{CD}  & = &
\rho  F^{\al\beta  }
 -\rho^2\xi(FGF)^{\al\beta} +\cdots ,\lb{yy1}\\
\{r^\al,p^\beta \}_{CD}  & = &
\delta^{\al \beta } - \rho\xi (GF)^{\al\beta}
+\cdots  .\lb{ry1}
\eeqa
Here $(GF)^{\al\beta}\equiv G^{\al\ga}F_\ga^\beta ,$ $(GFG)^{\al\beta}\equiv G^{\al\ga}F_{\ga
\s}G^{\s\beta}$ and we defined the field
strengths  as
\beqa
F_{\al \beta} & = &
\fr{\del \A_\beta}{\del r^\al}
-\fr{\del \A_\al}{\del r^\beta}
-\fr{i\rho }{\hbar} [\A_\al ,\A_\beta ] ,\label{fst}\\
G_{\al\beta}  & = &
\fr{\del \B_\beta}{\del p^\al}
-\fr{\del \B_\al}{\del p^\beta}
-\fr{i\xi }{\hbar} [\B_\al ,\B_\beta ] .
\eeqa
Moreover, we deal with the gauge fields satisfying  the conditions:
$ \B_\beta=  \B_\beta (p);\ \A_\al =\A_\al (r) ;\ [\A_\al ,\B_\beta ]=0  .$
Observe that
$r_\al$ and $p_\al$  should be considered as coordinates and the corresponding
momenta, respectively.
One can also show that the equations of motion of the phase space variables are
\beqa
{\dot r}^\al   =
 \xi \lp \fr{\del H_0}{\del r^\beta} -\fr{i\rho}{\hbar} [ \A_\beta  ,H_0]
\rp
\left(  G^{\al\beta  }
  -\rho\xi(GFG)^{\al\beta}+\cdots
\right) \no
  +\lp\fr{\del H_0}{\del p^\beta} -\fr{i\xi}{\hbar} [\B_\beta ,H_0] \rp
 \lp \delta^{\al \beta } - \rho\xi (GF)^{\al\beta}
+\cdots \rp , \lb{gss11}
\\
{\dot p}^\al  =
\lp\fr{\del H_0}{\del r^\beta} -\fr{i\rho}{\hbar} [\A_\beta  ,H_0]
\rp \lp-\delta^{\al \beta }
+ \rho \xi (FG)^{ \al\beta}  +\cdots \rp
\no
 +\rho \lp \fr{\del H_0}{\del p^\beta} -\fr{i\xi}{\hbar} [\B_\beta ,H_0] \rp
\lp   F^{\al\beta  }
 -\rho\xi(FGF)^{\al\beta} +\cdots
\rp  .\lb{gss21}
\eeqa

The method which we presented is  valid
for non--Abelian gauge fields which may be internal or external.
To clarify this point let the gauge field $\A $  be Lie algebra valued,  defined as
$\A_\al = \A_\al^n T_n.$ Here
$T_n$ are the generators of a Lie algebra
with the structure constants $f_{nmk}$:
\be
\label{lal}
[T_n,T_m]=f_{nmk}T_k.
\ee
Hence, the field strength components are given as
\be
\label{fla}
F^n_{\al \beta}  =
\fr{\del \A^n_\beta}{\del r^\al}
-\fr{\del \A^n_\al}{\del r^\beta}
-\fr{i\rho }{\hbar} f^{nmk}\A^m_\al \A^k_\beta  .
\ee
The difference between internal and external  gauge fields
resides in the $\hbar$ dependence of the coupling constant $\rho $\cite{gk}:
for an internal gauge field the coupling constant $\rho$
can be chosen to be linear in $\hbar ,$ but for  an external gauge
field  the coupling constant $\rho$ does not possess any $\hbar$
dependence. In the latter case the field strength (\ref{fla})
is singular in the $\hbar \rightarrow 0 $ limit.

\section{Dynamics  in noncommutative space}\label{dyn}

Inspecting the basic relations of the phase space variables
(\ref{rr1})--(\ref{ry1})
one can observe that
to introduce noncommutativity of the coordinates $r_\al$ one can set
$\xi =1$ and choose
the gauge field $\B_\al$ appropriately so that
 its field strength yields
\be
G_{\al\beta}=\theta_{\al\beta}.
\ee
The noncommutativity parameter  $\theta_{\al\beta}$ is constant and antisymmetric.
Let us deal with the systems where the canonical Hamiltonian satisfies
\be
\frac{\del H_0}{\del p_\al}=\frac{p_\al}{m};\ [ \B_\al  ,H_0]=0 .
\ee

By keeping the terms which are first order in $\theta_{\al\beta}$ and second order in $\rho$
in (\ref{rr1})--(\ref{ry1}),
one can observe that the basic relations between the phase space variables are
\beqa
\{r^\al,r^\beta \}_{CD}  & = &
  \theta^{\al\beta  } , \lb{rr}\\
\{p^\al,p^\beta \}_{CD}  & = &
\rho  F^{\al\beta  }   -\rho^2(F\tet F)^{\al\beta}  ,\lb{yy}\\
\{r^\al,p^\beta \}_{CD}  & = &
\delta^{\al \beta } - \rho (\tet F)^{\al\beta}   .\lb{ry}
\eeqa
The equations of motion of
the phase space variables  follow from (\ref{gss11}),(\ref{gss21}) as
\beqa
{\dot r}^\al &  = &
 \lp \fr{\del H_0}{\del r^\beta} -\fr{i\rho}{\hbar} [ \A_\beta  ,H_0]
\rp   \tet^{\al\beta  }
  +\frac{p_\beta}{m}
 \lp \delta^{\al \beta } - \rho (\tet F)^{\al\beta} \rp, \lb{gss1} \\
{\dot p^\al }& = &
\lp\fr{\del H_0}{\del r^\beta} -\fr{i\rho}{\hbar} [\A_\beta  ,H_0] \rp
\lp-\delta^{\al \beta }
+ \rho  (F\tet )^{ \al\beta} \rp
 +\rho  \fr{ p_\beta}{m}
\lp   F^{\al\beta  }
  -\rho (F\tet F)^{\al\beta}\rp  ,\lb{gss2}
\eeqa
by retaining
up to the first  and the second order terms, respectively,  in $\theta$ and $\rho .$
(\ref{gss1}) can be utilized to write the momentum in terms of the velocity $\dot r_\al$ as
\be
\label{ppp}
p^\al = m \lp \delta^{\al \beta } + \rho (\tet F)^{\al\beta} \rp {\dot r}_\beta
- \lp \fr{\del H_0}{\del r^\beta} -\fr{i\rho}{\hbar} [ \A_\beta  ,H_0]
\rp   \tet^{\al\beta  }.
\ee
Making use of (\ref{ppp}) one can derive
the force acting on the system by  keeping the terms at most linear in the velocity as
\beqa
m{\ddot r}_\al & = &
 -\lp \fr{\del H_0}{\del r^\beta} -\fr{i\rho}{\hbar} [ \A_\beta  ,H_0] \rp
 \lp \delta_{\al\beta} -\rho (\tet F )_{\al\beta} \rp
-\frac{im}{\hbar}\tet_{\al\beta} \left[
 \fr{\del H_0}{\del r^\beta} -\fr{i\rho}{\hbar} [ \A_\beta  ,H_0]
, H_0 \right] \no
&&
+ \frac{im\rho }{\hbar}\tet_{\al\beta  }
\lp \left[F_{\beta\ga}, H_0 \right]+
\left[ \fr{\del H_0}{\del r^\beta} -\fr{i\rho}{\hbar} [ \A_\beta  ,H_0] ,\A_\ga \right]\rp {\dot r}_\ga
\no
&&
+m\tet_{\al\beta  }\frac{\del}{\del r_\ga} \lp \fr{\del H_0}{\del r^\beta} -\fr{i\rho}{\hbar} [ \A_\beta  ,H_0]
\rp    {\dot r}_\ga
+\left( \rho F_{\al\beta} - \rho^2(\tet FF)_{\al\beta}\right){\dot r}_\beta . \label{fnc}
\eeqa
In the commutative limit the force becomes
\be
m{\ddot r}_\al |_{\tet =0} =  -\fr{\del H_0}{\del r^\al} +\fr{i\rho}{\hbar} [ \A_\al  ,H_0]
+\rho F_{\al\beta}{\dot r}_\beta .
\ee
This is in accord with  \cite{ana1}  where the relativistic case was
considered.

\subsection{The dynamics of dipoles in noncommutative coordinates}\label{ss}

Let us deal with a neutral elementary particle of spin--1/2 possessing
the  magnetic and electric  dipole moments
$\bm \mu =\mu \bm \s $ and $ \bm d =d \bm \s   ,$ which moves in the external
electric and magnetic fields $\bm E$ and $ \bm B.$ Here $\bm \s$ are
the Pauli  matrices and
$\mu =\hbar \gamma_M,\ d =\hbar \delta_D $ where $\gamma_M,\ \delta_D $
are constants.
In \cite{ana1} the dynamics of this system in ordinary (commutative) space  was studied in the low energy limit.
The  interaction Hamiltonian employed in \cite{ana1}
can also be derived in  terms of Foldy--Wouthuysen transformations from the
underlying relativistic equation\cite{prfn}, which can be  adapted to  our
approach by choosing  the canonical Hamiltonian as
\be
\label{hoa}
H_0=\frac{p^2}{2m}-\bm \mu \cdot\bm B -\bm d \cdot \bm E ,
\ee
and introducing the vector field
\be
\lb{dgf}
\bm \A =-\bm \mu \times \bm E + \bm d \times \bm B .
\ee
Obviously, time reversal symmetry is violated for a nonvanishing $\bm d .$
Plugging (\ref{hoa}) and (\ref{dgf}) into (\ref{fst}) with the coupling constant
$\rho =-1/c $ leads to the related field strength:
\beqa
F_{\al \beta } & = &
\ebsg \del_\al \left(-\mu_\s E_\ga +d_\s B_\ga \right)
-\easg \del_\bet \left(-\mu_\s E_\ga +d_\s B_\ga \right)  \no
&&-\fr{2}{\hbar c} \eabg \left(-\mu E_\ga +d B_\ga \right)
\left(\bm \mu \cdot\bm E +\bm d \cdot \bm B \right) .\label{fhoo}
\eeqa

We can obtain the noncommutative version of this system in the framework outlined above. In fact,
the full force acting on the particle in noncommutative space
can be derived
employing (\ref{hoa}) and (\ref{fhoo}) in (\ref{fnc}).
The velocity dependent terms of the force (\ref{fnc}) are related
to the quantum phases
which  is  the subject of the  next section, so that
here we only present the force  for $\dot{\bm r}=0:$
\beqa
f_\al(\tet)|_{\dot {\bm r} =0} &= &
\left( \fr{\del}{\del r_\beta} \left( \bm \mu \cdot\bm B +\bm d \cdot \bm E \right)
+\fr{2}{\hbar c}\ebsg \left( \bm \mu \times \bm B + \bm d \times \bm E\right)_\s
\left(-\mu E_\ga +d B_\ga \right)\right)
\left(\de_{\al\beta} +\fr{1}{c} (\tet F)_{\al \beta} \right) \no
&& -\fr{2m}{\hbar}\tet_{\al \beta}\left( \bm \mu \times \bm B + \bm d \times \bm
E\right)\cdot\fr{\del}{\del r_\beta}
\left(  \mu \bm B +d \bm E \right) \no
&& -\fr{4m}{\hbar^2 c}\tet_{\al \beta} \ebsg (\mu B_\s +d E_\s )(-\mu E_\ga +d B_\ga )\left(
\bm \mu \cdot \bm B + \bm d \cdot \bm E \right) \no
&& +\fr{4m}{\hbar^2 c}\tet_{\al \beta} \left( -\bm \mu \times \bm E + \bm d \times \bm
B\right)_\beta
\left(  \mu \bm B +d \bm E \right)\cdot\left(  \mu \bm B +d \bm E \right) . \label{frf}
\eeqa
For $\tet_{\al\beta}=0$ this coincides with the force given  in \cite{ana1}.

In particular, considering $d=0$ and  constant $\mu$
interacting  with  homogeneous external fields,  the force (\ref{frf}) yields
\be
\label{dr0}
\bm f(\tet)|_{\dot{\bm r}=0}  =  -\fr{2\mu}{\hbar c} (\bm \mu \times \bm B ) \times \bm E
- \fr{4m\mu^2}{\hbar^2 c} \left[(\bm \mu \cdot \bm B) \bm \tet \times (\bm B \times \bm E)
+ B^2  \bm \tet \times (\bm E \times \bm \mu )\right],
\ee
where we introduced
$
\tet_{\al \beta } = \epsilon_{\al \beta \ga} \tet_\ga
$
and neglected the terms of the order $1/c^{2} .$
One can observe that the force (\ref{dr0}) is nonvanishing when both external electric and magnetic
fields are present.
In the commutative limit  the force (\ref{dr0}) may be detected\cite{wr}.
When the experimental data
is available it may be used to set a bound on  $\tet$ employing the scheme which  will be discussed in the last section.

\section{The quantum phase from the (semi)classical force}\label{fvp}

We would like to derive the quantum phases from  the (semi)classical
forces  employing the relation given in \cite{spa}:
The integral
along a closed path $C$
of the classical force $f_\al$
acting on a particle of mass $m$ coupled to
the Abelian gauge field $a_\al$  with
the coupling constant $\eta$
leads to
\be
\label{fp}
\frac{i}{\hbar} \oint_C f_\al dr_\al =-i\frac{d\Phi}{dt},
\ee
where $\Phi$ is the quantum phase defined as
\be
\label{phi}
\Phi =\frac{\eta}{\hbar}\oint_C a_\al dr_\al .
\ee
This may seem to be a controversy because
generally  on the path where the quantum phase is calculated
there is no force acting on the particle. The force  can be considered to be
hypothetical as it was suggested in \cite{spa}. Nevertheless, it is an efficient procedure of obtaining the change in
the action when a vector field is coupled to the system, even if the quantum treatment becomes complicated (see also  \cite{hb}).
It was also shown in \cite{ana2} that  the quantum phases  and the  equations of motion which
lead to the force are interrelated.

When one deals with the non--Abelian gauge field $a_\al \equiv a_\al^n T_n$
where the generators of the Lie algebra $T_n$ satisfy the commutator (\ref{lal}),
the related phase factor can be given by the path ordering $P$
which  can be calculated as
$$
P \exp\left( \frac{-i\eta}{\hbar} \oint_{C_\e}  a_\al dr_\al \right)
= 1 -  \frac{i\eta}{2\hbar}f_{\al\beta}({\bm r}_\e ) s_\e^{\al\beta} ,
$$
for an infinitesimal closed curve  ${C_\e}$ around the point ${\bm r}_\e .$
Here $s_\e^{\al\beta}$
is the area of the infinitesimal surface spanned by  ${C_\e}$
and $f_{\al \beta} $ is  the related non--Abelian field strength.
Thus, for  non--Abelian gauge fields  we can generalize (\ref{fp}) as
\be
\label{fpna}
\frac{i}{\hbar} \oint_{C_\e} f_\al dr_\al =
-i\fr{d \Phi_\e}{dt},
\ee
where the infinitesimal phase is defined by
\be
\label{ifl}
\Phi_\e \equiv\frac{\eta}{2\hbar}f_{\al\beta}({\bm r}_\e )
s_\e^{\al\beta}.
\ee

Now, we would like to demonstrate that
in the ordinary space  our semiclassical approach is in accord with
this formulation.
Hence let us set $\B_\al =0$ and  $H_0=p^2/2m$.
Obviously, adding a  potential term like $V(r)$  to the free Hamiltonian does not affect
the quantum phase. The equations of motion (\ref{gss11}),(\ref{gss21}) give
\be
{\dot r}_\al =\fr{p_\al}{m},\  {\dot p}_\al =\rho F_{\al \beta} {\dot r}^\beta .
\ee
The integral of the force $m{\ddot r}_\al ={\dot p}_\al$
along the infinitesimal path $C_\e$ around the point ${\bm r}_\e $ yields
\be
\label{nana}
\rho \oint_{C_\e} F_{\al\beta} {\dot r}^\beta dr^\al =
\rho  F_{\al \beta}({\bm r}_\e ) \oint_{C_\e} \dot r^\beta dr^\al
= -\fr{\rho}{2}   F_{\al\beta} ({\bm r}_\e )
\fr{d s_\e^{\al\beta}}{dt}
= -\fr{\rho}{2} \fr{d}{dt}(  F_{\al\beta} ({\bm r}_\e )
s_\e^{\al\beta}),
\ee
where we suppose that the infinitesimal loop is moving so that
 $ds_\e^{\al\beta}/dt = \oint_{C_\e} (\dot r_\al dr_\beta- \dot r_\beta dr_\al )$
is the rate of change of the area of the surface spanned by the curve $C_\e ,$
and $F_{\al \beta}$  does not depend on time explicitly.
Therefore, by substituting   $\rho$ and $F_{\al\beta}$ with $\eta$ and $f_{\al\beta}$
the relation (\ref{fpna}) follows. Except the points
where $F_{\al\beta}$ is singular, one can extend the infinitesimal $\Phi_\e$
to an integral over a finite surface $S,$ whether the gauge fields $\A_\al$  is Abelian or non--Abelian.
However, when the gauge field $\bm \A$ is Abelian the infinitesimal phase  can
be written as
\be
\lb{aif}
\Phi_\e =\fr{\rho}{\hbar}\left(\bm \nabla \times \bm \A \right)\cdot {\bm s}_\e
\ee
where the infinitesimal  vector area is defined by
 ${\bm s}_\e \equiv \oint_{C_\e} \bm r \times d\bm r /2 .$
Extension of (\ref{aif}) to a finite surface $S$  leads to
\be
\lb{aff}
\Phi =\fr{\rho}{\hbar}\int_S\left(\bm \nabla \times \bm \A \right)\cdot d{\bm s}
=\fr{\rho}{\hbar}\oint_{C} \bm \A \cdot d\bm r ,
\ee
which is the quantum phase.

\section{The quantum phases in noncommutative coordinates}\label{ncp}
We would like to generalize  the procedure outlined above to define the quantum phases
in noncommutative coordinates . We adopt (\ref{fpna}) as the definition of the quantum phase for a given force:
we set $H_0=p^2/2m$ which is adequate for the
configurations
which we will consider, so that the force in noncommutative coordinates (\ref{fnc}) becomes
\be
\label{nccf}
 f_\al (\tet ) = \left( \rho F_{\al\beta} - \rho^2(\tet FF)_{\al\beta}\right){\dot r}^\beta .
\ee
Hence,  calculation of  the integral of the $\theta$--deformed force (\ref{nccf}) along the infinitesimal closed path $C_\e ,$
using the approach employed in the ordinary case (\ref{nana}),
\be
\label{fpnan}
\frac{i}{\hbar} \oint_{C_\e} f_\al (\tet )  dr_\al =
-i\fr{d \Phi_\e (\tet )}{dt},
\ee
yields the definition of
the $\tet$ deformed quantum phase as
\be
\lb{ncip}
\Phi_\e (\tet) =\fr{\rho}{2\hbar} \left( F_{\al\beta} - \rho(\tet FF)_{\al\beta}\right)
s^{\al\beta}_\e .
\ee
We can extend (\ref{ncip}) to an integral over a  finite surface $S,$ as far as
one does not encounter any
singularity of the
integrand:
\be
\lb{ncfp}
\Phi (\tet) =\fr{\rho}{2\hbar} \int_S \e^{\al\beta\ga} \left( F_{\al\beta} - \rho(\tet
FF)_{\al\beta}\right)
ds_{\ga} .
\ee
When we deal with an Abelian gauge field  (\ref{ncfp})
is the deformation of the quantum phase  (\ref{aff}) at the first order in $\tet .$
However, for a non--Abelian gauge field we should consider
(\ref{ncip}).
One of the important properties of the original quantum phases is their independence  of the velocity.
Our formulation of   $\theta$ deformed phases  respects this property.

\subsection{The AB phase}

Let an electron beam encircles a
solenoid placed at the origin whose symmetry axis is along $r_3.$ Thus,
we set $H_0=p^2/2m ,$
$\rho =e/c$ and the field strength
vanishes outside the solenoid but inside it is given with the constant magnetic field $B$ as
$F_{ij} =\eij B .$
Here $i,j=1,2.$
By letting the $r_1r_2$ coordinates
be noncommuting, $\tet_{ij} =\eij \tet ,$ one gets the $\tet$ deformed AB phase from (\ref{ncfp}) as
\be
\label{NAB}
\Phi_{AB}(\tet )
 =\fr{eBS}{\hbar c} \left(1+\frac{eB\tet}{c}\right)  =
\Phi_{AB} \left(1+\frac{eB\tet}{c}\right),
\ee
where $S$ is the area enclosed in the solenoid
and $\Phi_{AB}=\fr{eBS}{\hbar c}$ is the AB phase
related to the flux  $BS$ through the solenoid.
This is in accord with the result of \cite{ao} which was obtained by generalizing
the effective Hamiltonian obtained in a particular gauge.

\subsection{Dipoles interacting with electromagnetic fields}

We would like to obtain the $\tet$ deformed  quantum phases for the
dipoles interacting with external electric and magnetic fields whose
dynamics was discussed  before, (\ref{hoa})--(\ref{dr0}). We deal
with the configuration where $ \bm \mu \cdot \bm B=0;\ \bm \mu \cdot
\bm E =0;\ \bm d \cdot \bm B=0;\ \bm d \cdot \bm E =0. $
 Under these conditions the Hamiltonian (\ref{hoa})
yields $H_0=p^2/2m$ and the gauge field (\ref{dgf})
satisfies: $ [\A_\al ,\A_\beta ]=0. $ We further choose $ \bm \mu =\mu
\hat r_3;\ \bm d =d \hat r_3 $ and let there be no change in the
dipoles along the external fields: $ \bm E \cdot \bm \nabla \mu =0;\
\bm B \cdot \bm \nabla d =0. $ In this framework the field strength
(\ref{fhoo}) can be written as
\be
F_{ij}=\e_{ij} \left( -\mu  \bm
\nabla \cdot \bm E +d \bm \nabla \cdot \bm B\right) .
\ee
Let the
external electromagnetic fields be in the radial direction such that
$\bm \nabla \cdot \bm E$ and  $\bm \nabla \cdot \bm B$ vanish except
in the small regions around the third axes:
\be
 \bm \nabla \cdot \bm E  =   \fr{ \la_e}{s^\prime} ;\
 \bm \nabla \cdot \bm B  =   \fr{\la_m}{s^{\prime\prime}} . \label{db}
\ee
Here   $s^\prime $ and $s^{\prime\prime} $ indicate, respectively,  the  areas of the
regions
where
  $\bm \nabla \cdot \bm E$ and  $\bm \nabla \cdot \bm B$ do not vanish, though
usually one uses the limit values
\be
\label{lssp}
\lim_{s^\prime \rightarrow 0} \fr{1}{s^\prime } =
\lim_{s^{\prime\prime}  \rightarrow 0} \fr{1}{s^{\prime\prime} } =\delta (r_1)\delta (r_2).
\ee
$\la_e$ is the charge per unit length of the straight line charge distribution along the
third axis. Similarly,  $\la_m$ is related to magnetic monopoles or dipoles.
The $\tet =0$ part of the phase (\ref{ncfp}) yields
\be
\Phi_A =\fr{-1}{\hbar c}\int  F_{12} dr_1dr_2 = \frac{1}{\hbar c}
(\mu \la_e -d \la_m),
\ee
whether one uses
(\ref{db}) with
the  areas $s^\prime ,\ s^{\prime\prime} $ or their limits
(\ref{lssp}).
However, if one employs (\ref{lssp}) instead of
the  areas
$ s^\prime ,\ s^{\prime\prime} $  attributed
to the line of electric and magnetic
charges  in (\ref{db}),
the $\tet$ dependent part of the phase (\ref{ncfp}) will not be well defined.
Hence, we keep $ s^\prime ,\ s^{\prime\prime} $ and from (\ref{ncfp}) obtain the  Anandan phase
for the noncommutative $r_1r_2$ coordinates  as
\be
\label{PAC}
\Phi_A (\tet ) = \Phi_A \left( 1 + \fr{\tet \mu \la_e}{cs^\prime}
-\fr{\tet d \la_m}{cs^{\prime\prime}}  \right) .
\ee
Obviously, the AC and HMW phases
in noncommutative coordinates
follow for $d=0$ and $\mu =0$, respectively.

\subsection{Particle interacting with a  noncommutative gauge field}

In the non--Abelian physical  systems considered above the gauge fields $\A$ were internal.
However, one can also deal with external gauge fields.
When external non--Abelian gauge fields
are considered the  field strength (\ref{fla}) can be inserted into
(\ref{ncip}) to obtain the related quantum phase in noncommutative space.
Moreover, one can generalize this procedure  to the  noncommutative $U(N)$--valued
 gauge fields $\tilde{\A}_\al =\tilde{\A}_\al^nT_n$ by replacing the
ordinary product with the star product, so that the field strength is given by
\be
\label{fsnc}
\ti F_{\al \beta}  =
\fr{\del \ti \A_\beta}{\del r^\al}
-\fr{\del \ti \A_\al}{\del r^\beta}
-\fr{i\rho }{\hbar} \left(\ti \A_\al \star \ti \A_\beta -\ti \A_\beta \star \ti \A_\al
\right) .
\ee
The star product is defined in terms of another constant, antisymmetric  parameter
$\tilde{\tet}_{\al \beta}$ as
\be
\lb{star}
\star   =
\exp \left[
i \tilde{\tet}_{\al \beta}  \left(
\frac{\overleftarrow{\del}}{\del r^\al}
\frac{\overrightarrow{\del}}{\del r^\beta}
\right) \right] .
\ee
We distinguish the noncommutativity parameter of the space $\tet$ from
the noncommutative parameter of the gauge group $\tilde \tet ,$
because in quantum mechanical systems noncommutativity of space does not imply automatically
that the related gauge groups should be noncommutative.
Obviously, one can let the deformation parameters $\tet$ and $\tilde \tet$
to be the same by taking care of the approximations regarding $\tet$--dependent terms.

Let us deal with  the noncommutative $U(1)$ where
 the canonical Hamiltonian is   $H_0=p^2/2m$ and
the coupling constant is $\rho =e/c .$
By inserting (\ref{fsnc}) into (\ref{ncip}) one can obtain the deformed quantum phase for an
infinitesimal loop as
\be
\Phi_{NC} (\tet ) =\fr{e}{2\hbar c} \left( \ti F_{\al\beta}
 - \fr{e}{c}( \tet \ti F \ti F )_{\al\beta}\right) s_\e^{\al\beta} .
\ee
For  $\tet =0$ but $\ti \tet \neq 0$
this coincides with the quantum phase
obtained in \cite{clst}  considering
a gauge covariant generalization of the AB phase in noncommutative space.

\section{Discussions}

The most known examples of quantum systems where coordinates are effectively
noncommutative are   the charged particles moving on a plane in the presence of
a  strong perpendicular magnetic field
and the strings moving in the Neveu--Schwarz background field. In both of these systems
noncommutativity parameter  is proportional to the inverse  of  the field involved.
Thus for quantum mechanical  systems
in noncommutative coordinates  the value of the noncommutativity parameter
$\tet$ depends on the system considered. By adopting this interpretation
we would like to derive upper limits on the value of $\tet $
taking into account the quantum phases where experimental verifications are available.
However, before proceeding note that semiclassical brackets like  Poisson brackets,
possess a dimension of $\hbar ^{-1}.$ Hence,
if  we demand that  the dimension of the noncommutativity  parameter $\tet$
be $(\rm{length})^2,$  (\ref{rr}) dictates that we should perform the rescaling
\be
\label{rest}
\tet_{\al \beta } \rightarrow \fr{\tet_{\al \beta }}{\hbar} .
\ee

In \cite{cpst} the observed value  of the AB phase reported in  \cite{abe} was
used to derive an upper bound for
the noncommutativity parameter $\tet_{AB} .$  We adopt the same approach:
let $\Delta_{AB}$ denote percentage of the error in measuring the AB phase.
Comparing it with
the AB phase in noncommutative coordinates (\ref{NAB}) after the rescaling (\ref{rest}),  one  obtains the bound
\be
\tet_{AB} \leq \fr{\hbar c}{e B} \Delta_{AB} = \left( \fr{4.1}{2\pi}  \times 10^{-7}\ {\rm Gauss\ cm^2} \right)
\fr{\Delta_{AB}}{B}.
\ee
Hence,  the noncommutativity parameter regarding the AB effect $\tet_{AB}$ is proportional to the inverse of the
 magnetic field $B$.
Experimental error reported in \cite{abe} is $\Delta_{AB} = 20\%$ and the order of magnitude of the magnetic
field is $B\approx 10^4\ {\rm Gauss}.$
Therefore, we get the bound
\be
\label{labp}
\sqrt{\tet_{AB}} \leq 10^{-6} \ {\rm cm} .
\ee

By setting $d=0$ in (\ref{PAC}) one obtains the AC phase in noncommutative coordinates
after the rescaling (\ref{rest}) as
$$
\Phi_{AC} (\tet ) = \Phi_{AC} \left( 1 +\fr{\tet\mu \la_e}{\hbar c s^\prime }\right),
$$
where  $\Phi_{AC} =\fr{\mu \la_e}{ \hbar c}$ is the ordinary AC phase.
Now, in terms of the experimental error percentage $\Delta_{AC}$ we get the bound
on the noncommutativity parameter related to the AC phase as
\be
\tet_{AC} \leq \fr{ s^\prime }{ \Phi_{AC}}  \Delta_{AC}.
\ee

Considering neutrons, the AC phase was measured  in \cite{ace}  as
$$
\Phi_{AC}^{(n)} =2.19\pm 0.52 \ {\rm mrad}.
$$
If we let $s^\prime $ be of the order of the area enclosed within the
path of the neutrons given in \cite{mz} as  $s^\prime \approx 3 \ {\rm cm^2},$ we get
\be
\label{lacp}
\sqrt{\tet_{AC}^{(n)}} \leq 17 \ {\rm cm}.
\ee

The AC phase was measured  in \cite{AC2} dealing with a superposition of  two coherent beams of atoms possessing different magnetic
moments as
$$
\Phi_{AC}^{(a)} =2.22\pm 0.11 \ {\rm mrad}.
$$
Although, the geometry of the experiment is different from the originally  proposed one \cite{ac},  we  take $s^\prime \approx 0.8 \ {\rm cm^2},$
due to the fact that the beam travels in a region of diameter  $\approx 1\ {\rm cm}.$ Hence,  the bound reads
$$
\label{lacp1}
\sqrt{\tet_{AC}^{(a)}} \leq 4 \ {\rm cm}.
$$

The bounds which we obtained (\ref{labp}) and (\ref{lacp}) do not coincide with the ones derived
in \cite{cpst} and \cite{mz}, respectively. And this is because,
the quantum phases in noncommutative coordinates which we obtained
do not refer to energy of the particles in contrary to the ones attained in
the latter approaches.

\end{document}